# Beyond Technical Motives: Perceived User Behavior in Abandoning Wearable Health & Wellness Trackers

Ahmed Fadhil
University of Trento
Trento, Italy
ahmed.fadhil@unitn.it

## ABSTRACT
Health trackers are widely adopted to support individuals with daily health and wellness activity tracking. They can help increase steps taken, enhance sleeping pattern, improve healthy diet, and promote the overall health. Despite the growth in wearable adoption, their real-life use is still questionable. While some users derive long-term values from their trackers, others face barriers to integrate it into their daily routine. Studies have analysed technical aspects of these barriers. In this paper, we analyse the behavioural factors of discouragement and wearable abandonment strictly tied to user habits and lifestyle circumstances. A data analysis was conducted on 8 of the highly rated wearables for 2017. The analysis collected sale posts on Kijiji and Gumtree, the second sales online retailers for both the Italian and UK market, respectively. We extracted insights from the posts about user motives, highlighted technology condition and limitations, and timeframe before the abandonment. The findings revealed certain user behavioural patterns when abandoning their wearables. In addition, analysing the posts showed other motives for the posts and not strictly related to wearable abandonment.

## KEYWORDS
Health trackers; mobile healthcare; long-term engagement; wearable abandonment; behaviour change

## 1 INTRODUCTION

Wearable trackers and smartwatches have emerged as a way to track aspects of health and wellness. These trackers provide activity data about steps taken, sleeping patterns, diet tracking, and heart rate monitoring. Users expect to improve their health status by observing lifelong with these trackers through a new user experience [22]. Although user optimism about the prospects of wearables, there exists a gap in the wearable functionalities and user expectations and adoptions. Currently 30% of users stop using their tracker within 6 months [13]. According to Gartner's 2016 consumer survey [16], user's boredom of their wearables and smartwatches result in an abandonment rate of 29-30%. The survey stated that smartwatch adoption is still in early adopter stage (10%), while fitness trackers have reached early mainstream (19%). The study suggested that top reasons for abandonment include lack of usefulness, boredom, or the device malfunction. The study showed that people purchases smartwatches and fitness trackers for their own use, with 34% of fitness trackers and 26% of smartwatches received as gifts. Another study by Endeavour Partners in the US found that while one in 10 American adults own some form of activity tracker, half of them no longer use it. Hence, the wearable/smartwatch abandonment rate is higher relative to the usage rate. Previous studies analysed technical aspects behind the abandonment [5, 26], we took a step further and included also user behavioural aspects of abandonment. In this paper we reviewed posts on second sale fitness trackers and smartwatches on both Kijiji[1] and Gumtree[2], for both the Italian, and UK second sale wearable tracking technologies. The wearables searched for were among the best wearables for 2018 according to PCMAG[3] review, a Tech, Gaming, Healthcare & Shopping review site. We considered Apple Watch, Fitbit Charge 2, Fitbit Surge, Garmin Forerunner, Mi Band 2, Misfit, Samsung Gear, and TomTom Spark as the selected wearables and smartwatches for the study. In Table-1 we list the wearables and their given review details.

After wearable selection, we checked both sites for sale announcements about the selected wearables. A total of 484 posts about all wearables were reviewed from Kijiji and 624 posts from Gumtree. The review focused on analysing user motivation for post, health domain discussed within the post, adoption method and reasons for abandonment, usage frequency and duration before abandonment. We checked the wearable condition and technology limitations highlighted by users. All the analysed data were extracted from user posts and analysed to obtain accuracy with respect to user motivation and wearable features. Finally, we compared the findings from these studies with the study performed by Clawson et al., [5] on Craigslist[4] with a larger number of wearable trackers and fitness devices.

The findings revealed common motives for the abandonment, although the adoption motive was for different health domains, including physical activity, sleep pattern, and food tracking. In both of our analysis, users abandoned their wearables for motives other than perceived lack of utility or borne out of frustration and boredom with the device. For example, the analysis revealed abandonments for reasons related to double gifts received by users, upgrading to newer version or different models, or selling for wrong purchase. The study also revealed that majority of wearables featured at least three health and wellness elements, including physical activity, sleep and diet tracking. The frequency of use before abandonment varied from 1 day to few years. Several technical limitations were extracted from the posts on both datasets and highlighted as reasons for wearable abandonment. We consider these findings in our design recommendations for wearable technologies that focuses on user emotion and sustain evolving adoption. We will discuss particular recommendations based on behavioural theories, user motivation, design aesthetics, such that user experience and emotion are at the centre of the technology. Our work contributes to existing research on wearable technologies and user's engagement by analysing patterns in user engagement or frustration with the wearables. These will help advance the research on wearable tech- nologies and evolving adoption rates.

---
[1] https://www.kijiji.it/
[2] https://www.gumtree.com/
[3] https://www.pcmag.com/
[4] https://fortwayne.craigslist.org/

| Featured Fitness Trackers | | | |
|---|---|---|---|
| Smartwatches/Wearables | Ratings | Price | Bottom Line |
| *Apple Watch* | 4 - 5 | $369.00 (Apple Store) | The Apple Watch boasts sporty styling and some exclusive features that make it a solid alternative to the Series 2 smartwatch, especially for runners. |
| *Fitbit Charge 2* | 4.5 - 5 | $149.95 (Dell) | The Fitbit Charge 2 does everything the Fitbit Charge HR can, along with new idle alerts, automatic activity tracking, guided breathing sessions, interchangeable bands, and the option to connect your phone for GPS. |
| *Fitbit Surge* | 5 - 5 | $199.95 (Amazon) | With continuous heart rate monitoring, GPS, and broad appeal, the Fitbit Surge is the best all-day fitness tracker. |
| *Garmin Forerunner* | 4.5 - 5 | $449.99 (Amazon) | The Garmin Forerunner fitness tracker gives pertinent information to triathletes about their sports, including advice you don't often see, like recovery time. |
| *Mi Band 2* | 4 - 5 | $28 (Amazon) | The new version of Mi Band 2 has an OLED-display and a touch panel, which provides more information to its users. |
| *Misfit* | 4 - 5 | $39.99 (Amazon) | The Misfit combines top-notch fitness and sleep tracking with one of the best-looking designs. |
| *Samsung Gear* | 4 - 5 | $127.99 (Amazon) | With some solid improvements over its predecessor, the Samsung Gear Fit is a powerful fitness tracker and a fairly functional smartwatch. |
| *TomTom Spark* | 4 - 5 | $229.99 (Amazon) | The TomTom Spark Cardio + Music fitness tracker provides continuous heart rate monitoring, GPS with route tracking, excellent battery life, and music streaming, all wrapped up in a lightweight, waterproof design. |

**Table 1: The Selected Wearables and their Details.**

## 2 BACKGROUND
### 2.1 Wearables as Gadgets
Wearable fitness and smartwatch trackers are anticipated to proliferate in the market in the near future. Researchers have called the movement of tracking all aspects of one's daily life Life-logging [31], quantified self [4], or personal informatics [24]. Wearables are capable of tracking steps, and other physiological information (e.g., heartbeat rate). Wearables use data stored to allow users gauge progress and gather incremental feedback. The data are visualised to enhance user's awareness about everyday activities and facilitate independent living and improve quality of life for citizens [27]. Moreover, to motivate users keep up with their personal activities, some wearables adapt motivational techniques, such as gamification and social recognition [6, 34]. The design space to leverage tracking data and persuade health-related behaviour change involves issues and strategies to capture information, monitor progress, notify feedback, and provide social support [11, 19].

### 2.2 Wearables as IoTs
Wearables are often touted as the greatest applications of the Internet of Things. This technology has the potential to transform our lifestyle by tracking health and exercise progress and bringing smartphones power to the wearer's wrist. Wearable devices can absorb extremely rich source of contextual information, such as conversation, location, and gesture. A work by Billinghurst et al. [3] stated that wearables could sense handshakes, trigger face-recognition and identify individuals. Although their benefit in terms of low-cost personalised healthcare gadgets, however, previous literature have identified several (non) technical and design related issues acting as barriers for wearable adoption in long-term [5, 20]. A work by Lam et al. [20] implemented a wireless wearable biosensors platform that utilises biosensors and smartphone to measure heart rate, breathing rate, oxygen saturation, and estimate obstructive sleep apnea. The study addressed practical challenges in the design perspective of the platform. Another study by Jameson et al. [17] developed a mini wearable sensor device to enhance safety during ambulation for visually impaired users. The sensor warns them when they're about to hit obstacles at head level. The system emits an acoustic warning signals when a hazard is detected [17]. A similar work by Dakopoulos et al. [7] presented a survey among portable obstacle detection system as assistive technology for visually impaired people. The study analysed features and performance parameters of these wearables and provided a ranking as reference point of each wearable. Since wearables generate a vast amount of data, protecting this data is essential. A study by Pipada et al. [29] conducted a survey to gauge consumers concern about data security for wearable devices. The study explored possibility to accommodate these securities by the Technology Acceptance Model.

### 2.3 Wearables & Heterogeneous Demographics
Even though the wide adoption rate of fitness trackers and smartwatches, they have low user adoption rate for the long-term [32]. The majority of studies focused on technical aspects of wearable technologies and its correlation with user adoption [2, 35], little is known about individual characteristics and its correlation with wearable activity trackers adoption. Shih et al. [32] performed a six-week user study with 30 users using physical activity trackers embedded in clip-on and smartwatch physical devices. The study described user pattern implications, such as helping people be mindful of their physical activity tracker, to further articulate gender differences in use and adoption of wearable devices. To enhance user engagement in activity tracking, some wearables adapt exergames and gaming techniques. Wearable builders should engage users with incentives and gamification. Lindberg et al. [25] developed Running Othello 2 (RO2) exergame, in which players use a smartphone connected wrist band to compete in a board game enhanced with physical and pedagogical missions. The game uses inertial sensors and heart rate meter to detect physical activities of players. The findings revealed player's engagement with the game and identified challenges faced by users and how exergames with wearables could help.

## 2.4 User Rationale

Users are optimistic about the prospects of wearable devices, however, there is a gap in reliability, ease of use, user expectation, and interpretation of measured data from wearables [36]. Users often have different expectations for health management functions which is higher than that of daily auxiliary-type functions, which is an issue to be addressed by wearable manufacturers [36]. User preferences and rationale are among the strongest adoption and abandonment factors [5]. Successfully adopting such technology is an opportunity for new interventions to tackle health issues. However, existing wearables fail at many levels to sustain user engagement in the long-term. Clawson et al. [5] performed an analysis on ads of secondary sales of wearables on Craigslist[5]. They analysed 1600 ads of health tracking technologies announced over the course of a month. The study identified motives and rationale for wearable adoption and forsaking and suggested more work on behavioural theories to investigate psychological effects of health behaviour change techniques and the technologies that assist users make those changes. It is important to look at the behaviour of wearable users to understand the how and why they enjoy or abandon their device. A work by Lee et al. [22] presented experimental results on 80 college students, 64 days with fitness trackers. The study found that 50% of students have stopped wearing their tracker in less than three weeks at the first experiment and 34% have forsaken the device in nine weeks at the second experiment.

## 2.5 UX Design

Wearables still lack long-term user engagement due to factors, including wearable features, human perspective or behavioural aspects of the technology. A study by Lee et al. [23] discussed the sustainability aspect of wearables in improving individual's quality of life, social impact, and social public interest. The notion is that by recoding information about user behaviours, such as physical activity or diet, the wearable can educate and motivate these towards better habits and health. The gap of recording information and changing behaviour is substantial, however little evidence suggests that they are bridging that gap [23]. To create a personalised experience in wearable technologies, user experience with the device should be considered in early stages of product development. This will help explore good, bad and ugly features and factors for wearable technologies. Williams et al. [37] conducted two participatory design phases with a team of 8 visually impaired adults to explore new features for a new wearable navigation technology. The study compared low and medium fidelity prototype activities using office supplies and electronic components. The result revealed higher participant engagement during the medium fidelity session [37]. Although the role of wearable devices in facilitating health behaviour change, this change might not be driven by the wearable alone. Instead, Patel et al. [28] discusses that successful use and potential health benefits related to wearables depends on the design of the engagement strategies than on features of their technology. The engagement strategies, such as social competition and collaboration, and effective feedback loops are the one that connect with human behaviour [28].

[5] https://fortwayne.craigslist.org/

## 3 METHOD

To analyse technical and behavioural reasons behind wearable abandonment, we performed a thorough review on available announcements on Kijiji for the Italian market and Gumtree for the UK market for second hand sales of wearables. The study considered 8 of the highly ranked wearable activity trackers and smartwatches in the market, namely Apple Watch, Fitbit Charge 2, Fitbit Surge, Garmin Forerunner, Mi Band 2, Misfit, Samsung Gear, and TomTom Spark. We reviewed technical rationales behind the abandonment, however, the research focused on the behavioural aspect of the abandonment. This was because user engagement and motivation are strongly bound with behavioural aspects. We analysed the motives, the health domain, the adoption and abandonment reasons, the frequency and duration of use and abandonment, and the technical limitations mentioned in the posts.

### 3.1 Wearable Selection

We initially checked the best reviewed and highly rated wearables/smartwatches for 2018 on PCMAG technology review site. We picked wearables and smartwatches rated at least 4.5 - 5 stars, and strictly related to tracking aspects of health, such as diet, food journaling, sleeping patterns, physical activity and other health features, such as water and caffein tracking.

### 3.2 Announcement Collection

We initially assumed that when a person posts wearable, smartwatches for sell, they are abandoning the technology and trying to sell it for no use/no benefits. Nonetheless, our investigations of the announcements and users motives behind their abandonment revealed other rationale behind the announcement, and not strictly related to abandonment. We reviewed the posts found on Kijiji between 10 and 31 January 2018. We later checked for the sample trackers on Gumtree between 15 and 28 of August. We searched for each selected wearable and smartwatch and noted the associated posts returned for each website separately. We then conducted technical and behavioural analysis of the posts. The total number of announcements obtained was 158 for Apple Watch, 9 for Fitbit Charge 2, 15 for Fitbit Surge, 70 for Garmin Forerunner, 10 for Mi Band 2, 7 for Misfit, 208 for Samsung Gear, 7 for TomTom Spark, and 484 in total for the Kijiji. Whereas, it was 196 for Apple Watch, 116 for Fitbit Charge 2, 81 for Fitbit Surge, 80 for Garmin Forerunner, 12 for Mi Band 2, 39 for Misfit, 99 for Samsung Gear, 1 for TomTom Spark, and 624 in total for the Gumtree.

### 3.3 Data Analysis

The data analysis focused on motivations for the announcement, the health domains discussed in the posts, the adoption method and reasons for the abandonment, the frequency and duration of use before abandonment, and the technical limitations highlighted by users together with technology condition. In Figure-1 we show the pipeline followed to obtain and review the posts dataset with all the terms and features collected and deemed relevant to the posts on both Kijiji and Gumtree sites.

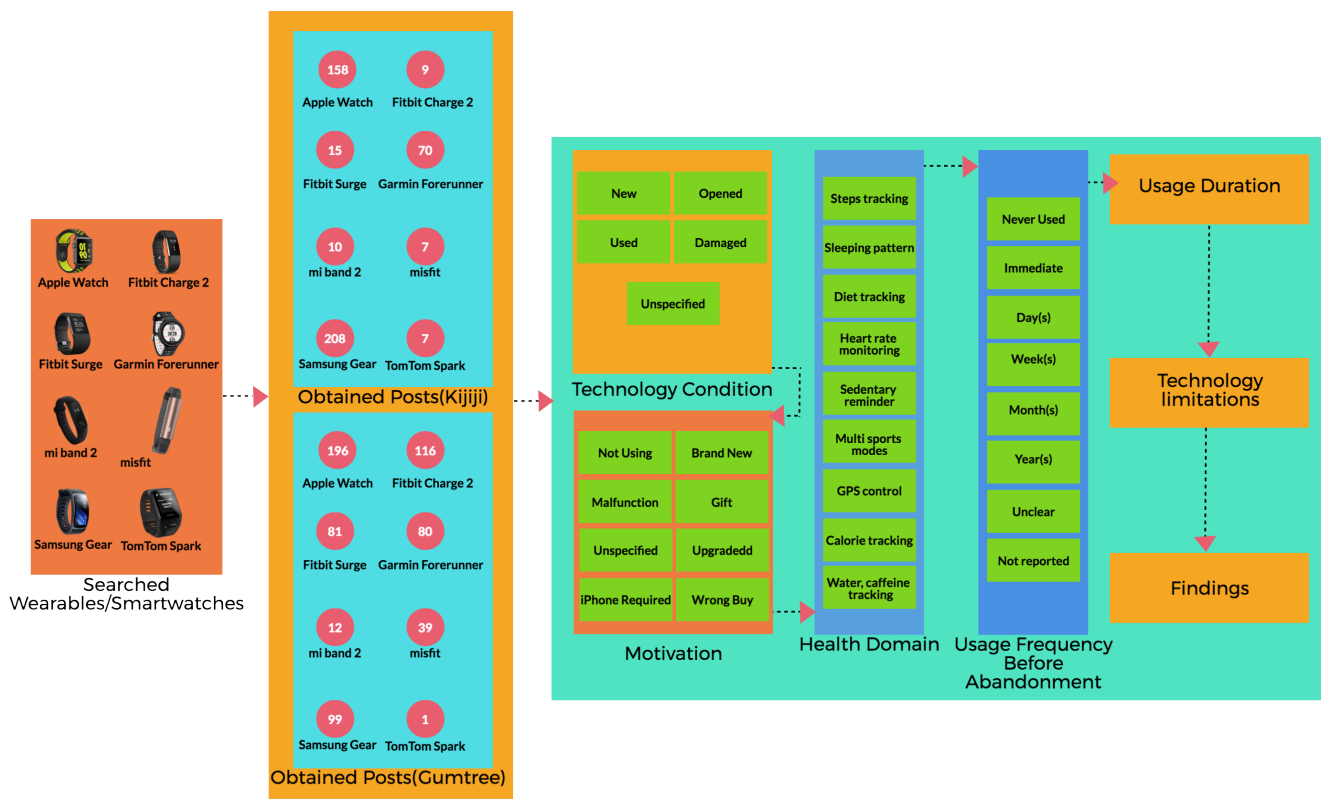

Figure 1: The Dataset Pipeline to Analyse Kijiji and Gumtree Posts.

## 3.4 Findings

Based on the review there exists patterns in user behaviour towards the announcement and reasons beyond the abandonment of the technology. For example, a motivation to sell a wearable was because of double gifts, which often times resulted in having two items of the same brand. This is not strictly a reason for abandonment, but rather most probably a continues use. Since the user didn't mention any intention to abandon the technology, but rather they had other motives behind their decision. Other trends in the data were related to the health domain covered by the technology and health reasons mentioned by users for the abandonment. For example, some decided to post their wearable because they had a health status change, as in *"I sell because I can't use it due to my heart problem"*, or because of a change in their lifestyle, as in *"I lost weight and achieved my goal, I don't use it anymore"* and *"I bought a bike and I can't use it to track my steps"*. The usage duration was spread and ranged from 1 day, as in *"Bought at the day one, its new"* to few years, as in *"2 years of life, but taken good care of"*. There were many similarity patterns among the posts obtained from both Kijiji and Gumtree. Interestingly, some (non) technical were obtained from the posts provided insights on why they abandon the wearable and what are the associated limitations led to the posts that were highlighted by the users. For example, some reasons were related to the bracelet size for the wearable, as in *"The band is too small for me"*, or for a feature requirement, as in *"The band is not water proof"*, or simply for user's personal preferences, such as colour, as in *"I sell because I bought a black one"*.

*3.4.1 Technology Condition.* The wearables and smartwatches found in the posts were of different conditions. The conditions obtained from the post were new, opened, used, damaged and some were unspecified for both datasets. The majority of the announcements posted that the technology was new, never opened, never used, or unused. Most of the sellers on Kijiji posted selling Samsung Gear (141 posts), and Apple Watch (49 posts). Whereas the posts on Gumtree were mostly about Apple Watch (196 posts) and Fitbit Charge 2 (116 posts). We obtained a variety of post information from Kijiji, whereas Gumtree had fewer post motives, but higher number of posts in the existing motives. For example, Kijiji had posts about iPhone Required and Wrong Buy, where sellers mentioned relevant motives for their sell. Interestingly we found some posts that mentioned the wearable was damaged, not working, or malfunction, although they were few (Apple Watch=3, Garmin Forerunner=2, Samsung Gear=1). We didn't find any posts discussing the above malfunction conditions in Gumtree as motives for their sell. In Table-2 we list the wearables/smartwatches and their condition obtained from the announcements.

*3.4.2 Motivation.* After the announcement search, we began by looking at motives for post announcements. We obtained 484 posts for Kijiji and 624 posts for Gumtree from all the 8 categories.

Beyond Technical Motives: Perceived User Behavior in Abandoning Wearable Health & Wellness Trackers

| Technology Condition | | | | | | | | |
|---|---|---|---|---|---|---|---|---|
| Condition | Apple Watch | Fitbit Charge 2 | Fitbit Surge | Garmin Forerunner | Mi band 2 | Misfit | Samsung Gear | TomTom Spark |
| **Kijiji Posts** | | | | | | | | |
| New(never opened)(never used) (unused) | 49 | 6 | 6 | 16 | 4 | 5 | 141 | 4 |
| Opened | 5 | 1 | 0 | 2 | 0 | 0 | 3 | 1 |
| Used | 59 | 2 | 0 | 32 | 1 | 2 | 48 | 2 |
| Damaged (not working) (malfunction) | 3 | 0 | 0 | 2 | 0 | 0 | 1 | 0 |
| Unspecified | 88 | 0 | 9 | 18 | 5 | 0 | 14 | 0 |
| **Gumtree Posts** | | | | | | | | |
| New(never opened)(never used) (unused) | 28 | 36 | 33 | 17 | 1 | 2 | 68 | 0 |
| Opened | 14 | 10 | 2 | 4 | 0 | 0 | 1 | 1 |
| Used | 20 | 15 | 6 | 8 | 1 | 0 | 0 | 9 |
| Damaged (not working) (malfunction) | 0 | 0 | 0 | 0 | 0 | 0 | 0 | 0 |
| Unspecified | 141 | 59 | 39 | 52 | 15 | 36 | 3 | 0 |

**Table 2: The Condition of Smartwatches/Wearables in Posts.**

There were more variety of motives for posting obtained from Kijiji than Gumtree. These motivations of the posts were mainly not using, brand new, malfunction, gift, unspecified, upgraded, iPhone acquired, and wrong buy. Whereas, the same motives found for Gumtree, however no post discussed malfunction, iPhone requirements and wrong purchase as their motives for sell. Most posts on Kijiji mentioned selling brand new items, from all the categories, with 171 posted items selling brand new wearables/smartwatches. This was followed by 34 posted items to sell because they claimed they received it as a gift. We obtained 25 posted items sold for no use. The posts on Gumtree have shown similar results as motives for sell. There were 185 posts selling brand new wearables/smartwatches. This was followed by 53 posted items to sell because they claimed they received it as a gift. Finally, we obtained 32 posted items sold for no use. We obtained 237 and 345 unspecified posts from both Kijiji and Gumtree, respectively. Although we didn't consider the unspecified posts among the motives for posting items, however understanding user intention behind these posts deems relevant to analyse their adoption and abandonment motives. The majority of announcements on Kijiji were about Samsung Gear with 208 posts, followed by Apple Watch, with 196 posts, followed by Fitbit Charge 2, with 116 posts. Although there were more posts about Apple Watch on both categories, this still isn't enough to conclude that they are among the most abandoned wearables/smartwatches. We believe the fact that this brand owns a big share in the wearable/smartwatch market [30] and having more posts about them could be because of the heigh number of adopters and not necessarily due to technical limitations or user frustration. In Table-3 we list the wearables/smartwatches and the user motivation for the post as was given by the announcements from both websites.

*3.4.3 Health Domain.* Most posts from both datasets mentioned one or more health domain the technology supports. The health domain mentioned on Kijiji were steps tracking, sleeping pattern, diet tracking, heart rate monitoring, sedentary reminder, multi sports modes, GPS control, calorie tracking, water and caffeine intake tracking. Whereas, the health domains on Gumtree were missing the diet tracking, sedentary reminder, and water, caffeine intake tracking. The review revealed that most health functionalities were mentioned as a way to promote the tracker and highlight the unique features they encompass. The Kijiji posts mentioned Fitbit Surge and Mi band 2 more often in terms of health and wellness features they provide. For instance, most of the posts about Fitbit Surge mentioned its steps, sleep and diet tracking functionalities. Whereas, Apple Watch was more frequent with Gumtree. In Table-4, we list the wearables/smartwatches and the health domain as given by the announcements on both data.

*3.4.4 Adoption Method.* This refers to the way in which the users have had the wearables/smartwatches. Most Kijiji posts mentioned the direct purchase and gifts reviewed as the main adoption method, the data was the same about Gumtree, except that we found one post mentioned winning the gadget. However, a big portion of the posts from both data didn't mention clearly the adoption mechanism. In Table-5, we highlight the adoption methods as provided by the announcements.

*3.4.5 Abandonment Reasons.* To investigate technological use, we had to check the abandonment reasons. For that, we analysed the reviewed posts from technical and nontechnical stand point to check for reasons behind their abandonment decision. Most of the abandonments on Kijiji posts were due to change in accessories, upgrading model, gift, user preferences, perceived usefulness, wrong purchase, tech curiosity, double items, compatibility issues, feature requirements and size issues. Whereas, Gumtree posts didn't reveal any abandonment due to wrong purchase, tech curiosity, size or compatibility issues. Although abandoning the technology is perceived negatively, however we found cases where the user was selling their device for upgrade or double gift, in which case its not considered as abandonment. In such cases the seller is still in the context of use, although he is switching or selling extra devices. In Table-6, we highlight reasons behind technology abandonment as was given by the announcements from both datasets.

*3.4.6 Usage Frequency.* This refers to the number of times the user at least opened and used the technology before abandoning it or deciding to post it online. Based on the announcements, we categorised the usage frequency into: never used, immediate, day(s), week(s), month(s), year(s), unclear, and not reported. These conditions were decided based on the extracted information from the posts. The categorisation is defined as follows:

- Never used: In which the item was received but never opened, never tested, never used at all.

| | Motivation | | | | | | | | |
|---|---|---|---|---|---|---|---|---|---|
| **Smartwatches/Wearables** | *Not Using* | *Brand New* | *Malfunction* | *Gift* | *Unspecified* | *Upgraded* | *iPhone Required* | *Wrong Buy* | **Total** |
| | **Kijiji Posts** | | | | | | | | |
| *Apple Watch* | 0 | 35 | 0 | 4 | 104 | 9 | 3 | 1 | **158** |
| *Fitbit Charge 2* | 0 | 2 | 0 | 1 | 6 | 0 | 0 | 0 | **9** |
| *Fitbit Surge* | 2 | 3 | 0 | 0 | 11 | 0 | 0 | 0 | **15** |
| *Garmin Forerunner* | 6 | 12 | 3 | 4 | 35 | 9 | 0 | 0 | **70** |
| *Mi band 2* | 1 | 4 | 0 | 1 | 4 | 0 | 0 | 0 | **10** |
| *Misfit* | 0 | 2 | 0 | 1 | 4 | 0 | 0 | 0 | **7** |
| *Samsung Gear* | 15 | 110 | 3 | 22 | 70 | 0 | 0 | 0 | **208** |
| *TomTom Spark* | 1 | 3 | 0 | 1 | 3 | 0 | 0 | 0 | **7** |
| **Total** | **25** | **171** | **6** | **34** | **237** | **18** | **3** | **1** | **484** |
| | **Gumtree Posts** | | | | | | | | |
| *Apple Watch* | 14 | 28 | 0 | 12 | 141 | 1 | 0 | 0 | **196** |
| *Fitbit Charge 2* | 10 | 36 | 0 | 10 | 59 | 1 | 0 | 0 | **116** |
| *Fitbit Surge* | 2 | 33 | 0 | 6 | 39 | 1 | 0 | 0 | **81** |
| *Garmin Forerunner* | 4 | 17 | 0 | 0 | 52 | 7 | 0 | 0 | **80** |
| *Mi band 2* | 0 | 1 | 0 | 0 | 15 | 0 | 0 | 0 | **12** |
| *Misfit* | 0 | 2 | 0 | 1 | 36 | 0 | 0 | 0 | **39** |
| *Samsung Gear* | 1 | 68 | 0 | 24 | 3 | 3 | 0 | 0 | **99** |
| *TomTom Spark* | 1 | 0 | 0 | 0 | 0 | 0 | 0 | 0 | **1** |
| **Total** | **32** | **185** | **0** | **53** | **345** | **13** | **0** | **0** | **624** |

**Table 3: Smartwatches/Wearables and User Motivation for Posts.**

| | Health Domain | | | | | | | | |
|---|---|---|---|---|---|---|---|---|---|
| **Smartwatches/Wearables** | *Steps tracking* | *Sleeping pattern* | *Diet tracking* | *Heart rate monitoring* | *Sedentary reminder* | *Multi sports modes* | *GPS control* | *Calorie tracking* | *Water, caffeine intake tracking* |
| | **Kijiji Posts** | | | | | | | | |
| *Apple Watch* | x | x | - | x | x | - | x | - | - |
| *Fitbit Charge 2* | x | | | x | - | x | - | - | - |
| *Fitbit Surge* | x | x | x | - | x | x | x | - | - |
| *Garmin Forerunner* | x | - | - | x | - | x | - | - | - |
| *Mi band 2* | x | x | - | x | x | x | | x | - |
| *Misfit* | x | x | x | - | x | - | - | x | - |
| *Samsung Gear* | x | - | - | x | - | x | - | - | x |
| *TomTom Spark* | x | - | - | x | - | x | x | - | - |
| | **Gumtree Posts** | | | | | | | | |
| *Apple Watch* | - | - | - | - | - | - | x | - | - |
| *Fitbit Charge 2* | x | x | - | x | - | - | - | x | - |
| *Fitbit Surge* | x | x | - | x | - | x | x | x | - |
| *Garmin Forerunner* | x | - | - | x | - | x | x | x | - |
| *Mi band 2* | x | - | - | x | - | - | - | x | - |
| *Misfit* | x | x | - | - | - | - | - | x | - |
| *Samsung Gear* | x | x | - | x | - | - | x | - | - |
| *TomTom Spark* | - | - | - | - | - | - | x | - | - |

**Table 4: Smartwatches/Wearables and Health Domain Discussion by Posts.**

| Adoption Method | | |
|---|---|---|
| **Kijiji Posts** | | |
| Gift (34 posts) | Purchased (213 posts) | Unspecified (237 posts) |
| **Gumtree Posts** | | |
| Gift (53 posts) | Purchased (110 posts) | Unspecified (345 posts) |
| Wining (1 post) | | |

**Table 5: Adoption Method.**

| Abandonment Reason | | |
|---|---|---|
| **Kijiji Posts** | | |
| Accessory change | Immediate | Upgrading model |
| User preferences | Perceived usefulness | Wrong purchase |
| Double items | Compatibility issues | Feature requirements |
| Tech curiosity | Size issues | Gift |
| **Gumtree Posts** | | |
| Accessory change | Immediate | Upgrading model |
| User preferences | Perceived usefulness | Double items |
| Feature requirements | Gift | |

**Table 6: Abandonment Reasons.**

- Immediate: In which an item was received, and opened for testing but never worn, not even for 1 day.
- Day(s): In which an item was received, opened and worn for one day only, or a few days, but less than a week.

- Week(s): In which an item was received, opened and worn for one week only, or a few weeks, but less than a month.
- Month(s): In which an item was received, opened and worn for one month only, or a few months, but less than a year.
- Year(s): In which an item was received, opened and worn for one year, or a few years.
- Unclear: Where the user mentions the usage but is not clear about the timeframe of use.
- Not reported: Where the user provided no information about the usage frequency or time.

Based on our analysis, most Kijiji posts relevant to Apple Watch and Samsung Gear had items that were never used, with Apple Watch=27 posts and Samsung Gear=110 posts. However, there were fewer Gumtree posts relevant to Apple Watch, and most new item posts were about Samsung Gear, 68 posts and Fitbit Charge 2, 36 posts. Apple Watch and Samsung Gear were also the highest in terms of immediate abandonment, with Apple Watch=14 posts and Samsung Gear=21. Whereas Fitbit Charge 2 was the highest in Gumtree in terms of immediate abandonment, with 25 posts. In Table-7, we highlight the frequency condition of technology use before the abandonment as was extracted from the announcements from both websites. There were 169 posts from Kijiji and 41 posts from Gumtree obtained from all the wearables/smartwatches that had no indication about the usage frequency, based on the post analysis.

*3.4.7 Usage Duration.* With duration we measured the period each technology was used, as given by the posts. This was helpful to predict from a given data the timeframe in which a user decides to abandon a given technology. Most of posts about Apple watch from both data had a usage duration from days, weeks, months, to years. This was followed by Samsung Gear in Kijiji data and Fitbit Charge 2 in Gumtree. Where posts mentioned a usage duration from days to months. None of the posts mentioned the duration of use for TomTom Spark. These timeframes were decided after reviewing information on the posts, where the user mentions the duration, they had the technology before they posted it for sell. Some examples were, *"I used it for 1 day just to test it"* for a daily duration, *"I used it for only one week"* for a weekly duration, *"I used it for 2 months only"* for monthly duration and *"It works perfectly it has a year of life"* for yearly durations.

## 3.5 Wearable Limitation

We reviewed the announcements and extracted wearables and smartwatches related limitations that were highlighted by users as reasons for the sell announcement from both Kijiji and Gumtree posts. We considered these limitations as reasons for abandonment. The limitations highlighted were sometimes technical, but often non-technical and were mainly related to user preferences. We constructed a list of keywords from the analysis that better describes what users have highlighted. These keywords were colour change, upgrading model, gift, user preferences, perceived usefulness, wrong purchase, tech curiosity, double items, compatibility issues, feature requirements, and size issues. Based on these keywords, we obtained the frequency of each limitations as given by the posts.

In the case of Kijiji posts, we found that the majority of users perceived the technology as useless (38 posts). Interestingly, this was followed by upgrading the model (19 posts). We believe that uselessness and upgrading aspects of limitations are very different, since it is very different from the uselessness aspect; the first one is resulting in abandoning the technology, whereas the second one is continuing the use. The rest of limitations found to be $\leq 6$ posts. Moreover, Samsung Gear was the most discussed technology in all the keywords, followed by Apple Watch.

In the case of Gumtree posts, the majority of users mentioned Gift, user preferences as the main reason for the post (14 posts). This was followed by Feature requirements with 11 posts, and Upgrading model with 8 posts, then Double items, where users stated the selling reason due to having two items of the same gadget. The rest of limitations found to be 3 posts. Moreover, Fitbit Charge 2 was the most discussed technology in all the keywords, followed by Garmin Forerunner.

The announcements mostly mentioned perceived uselessness of the technology. For example, a user mentioned *"I sell it because I can not use it"*. However, other cases included selling the old model to upgrade to newer models. For example, *"I sell by switching to 2 'series because of the need for GPS"*. Other less frequent cases included colour change (e.g.,*"I sell because I bought a black one"*); gift, user preferences (e.g.,*"I sell because it's a bad gift"*); wrong purchase (e.g.,*"I bought for incorrect steel bracelet purchase"*); tech curiosity (e.g.,*"Purchased for curiosity"*); double items (e.g., *"because I already have one like that and that's enough"*); compatibility issues (e.g., *"I sell because it is not compatible with my notes"*); feature requirements (e.g.,*"The band is not water proof"*); and size issues (e.g., *"The band is too small for me"*), (see Table-8 for technology limitations as perceived by the users from both post data).

## 4 FUTURE INSIGHTS

Understanding reasons behind user abandonment of wearables and smartwatches by investigating real posts is an important step towards finding insights about technical limitations led to the abandonment. However, focusing only on technical limitations is still not enough to fully understand user motives for the abandonment. Hence, in this study we took the investigation a step further and performed analysis on user posts about these technologies. Besides analysing technical limitations that led to the announcements, we particularly focused on user behaviour and intention for the abandonment through a thorough analysis of all the posts. We extracted intends and contextual meanings from the posts. The findings revealed some correlation pattern between wearable analysis from technical perspective and users' tone from behavioural perspective. Our goal was to provide insights on how to build enticing wearables, and to enhance their long-term use. The findings have led us to some future insights/recommendations to consider in the context of wearable abandonment. These insights answer our question: **How to increase wearable/smartwatch devices adoption rate?**.

*4.0.1 Usage Flexibility.* Successful trackers are those designed to solve specific problems [21]. However, wearables are yet another piece of exercise equipment, and they're easier to put away than an elliptical machine. Perhaps, wearables work better for those dedicated to maintain healthy lifestyle in long-term. The challenge is in engaging those who lose interest in the activity and abandon the device in the long-term. The device should be flexible in terms of activities it tracks and should provide services beyond data in and out approach. The device should give users a compelling reason to continue using it. For wearables to enhance long-term engagement and provide flexibility in tracking user changes, it has to have three factors: habit formation, social motivation and goal reinforcement [21]. The best engagement strategy to help habit formation process more effectively is to move beyond data presentation (e.g., steps, calories, stairs) and directly address elements of habit loop (e.g., cue, behaviour, reward) and trigger the sequence that resulted in new, positive habits. If users are effectively motivated, then they will continue wearing the device. Goal-reinforcement can help leverage user motivation, especially when they feel progress towards a defined goal. All these criteria help with building a flexible wearable, that considers not only technical aspects, but also the behavioural aspects and user intention to use the technology. Users interests evolves over time, and so should wearables, in parallel with users and adapt with new things and track new activities, such as jumping rope.

| Usage Frequency Before Abandonment | | | | | | | | |
|---|---|---|---|---|---|---|---|---|
| Smartwatches/Wearables | Never Used | Immediate | Day(s) | Week(s) | Month(s) | Year(s) | Unclear | Not reported |
| Kijiji Posts | | | | | | | | |
| Apple Watch | 27 | 14 | 3 | 2 | 6 | 3 | 53 | 72 |
| Fitbit Charge 2 | 2 | 2 | 1 | 0 | 1 | 0 | 1 | 2 |
| Fitbit Surge | 5 | 4 | 0 | 0 | 0 | 0 | 2 | 4 |
| Garmin Forerunner | 9 | 8 | 0 | 0 | 3 | 3 | 6 | 32 |
| Mi band 2 | 3 | 1 | 0 | 0 | 2 | 0 | 0 | 4 |
| Misfit | 3 | 1 | 0 | 0 | 0 | 0 | 0 | 3 |
| Samsung Gear | 110 | 21 | 0 | 0 | 4 | 0 | 11 | 49 |
| TomTom Spark | 1 | 1 | 0 | 0 | 0 | 0 | 2 | 3 |
| Gumtree Posts | | | | | | | | |
| Apple Watch | 2 | 8 | 5 | 5 | 0 | 0 | 0 | 9 |
| Fitbit Charge 2 | 110 | 25 | 8 | 0 | 0 | 0 | 0 | 3 |
| Fitbit Surge | 20 | 0 | 12 | 25 | 6 | 0 | 0 | 8 |
| Garmin Forerunner | 30 | 10 | 11 | 11 | 7 | 2 | 0 | 0 |
| Mi band 2 | 0 | 0 | 0 | 5 | 8 | 11 | 0 | 4 |
| Misfit | 3 | 3 | 2 | 0 | 0 | 0 | 0 | 8 |
| Samsung Gear | 0 | 0 | 5 | 12 | 6 | 3 | 0 | 7 |
| TomTom Spark | 23 | 8 | 4 | 0 | 0 | 0 | 0 | 2 |

**Table 7: Usage Frequency Before Abandonment.**

| Technology limitations | | | | | | | | | |
|---|---|---|---|---|---|---|---|---|---|
| Keywords | Frequency | Apple Watch | Fitbit Charge 2 | Fitbit Surge | Garmin Forerunner | Mi band 2 | Misfit | Samsung Gear | TomTom Spark |
| Kijiji Posts | | | | | | | | | |
| Color change | 3 | x | | | | | | | |
| Upgrading model | 19 | x | | | x | | | x | |
| Gift, user preferences | 6 | x | | | | | x | x | |
| Usefulness | 38 | x | x | x | x | x | | x | x |
| Wrong purchase | 5 | x | | | | | | x | x |
| Tech Curiosity | 2 | | x | | | | | x | |
| Double items | 4 | | | | | | | x | |
| Compatibility issues | 2 | | | | | | | x | |
| Feature requirements | 2 | | | x | | | | | |
| Size issues | 1 | | | | | | | x | |
| Gumtree Posts | | | | | | | | | |
| Color change | 1 | x | x | | | | | | |
| Upgrading model | 8 | | | x | | | | | x |
| Gift, user preferences | 14 | | x | | | | x | x | |
| Usefulness | 3 | | x | | x | | | | |
| Wrong purchase | 1 | x | x | | x | | x | | x |
| Tech Curiosity | 2 | | x | | | | x | x | |
| Double items | 7 | x | x | | x | | | x | |
| Compatibility issues | 1 | | | | x | | | | |
| Feature requirements | 11 | | x | | x | x | | x | |
| Size issues | 1 | | x | | | | | | |

**Table 8: Technology limitations Perceived by Users.**

*4.0.2 Design Aesthetics.* Building the right user interaction design and experience is crucial in improving long-term engagement with the wearable. The design aesthetics greatly decides whether a user continues to use the wearable or abandon it after few tries [10]. According to ledger et al. [21] even with a great product, if its missing design aesthetics, then it is hard to expect users adopting it in the long-term. Among the design insights found in this study were products that fit well, comfortable and compatible with lifestyle. For example, one post complained about the product not being waterproof, and hence decided to sell it. This is essential feature if it requires being worn for 24/7, so to obtain accurate data. Among design flaws current wearables suffer from not being waterproof, their insufficient battery life and being uncomfortable to wear due to size or other user preferences [10]. Any of these flaws can lead to turn off user engagement with the device in the long-term.

*4.0.3 Theoretical Foundation.* Although the importance of design aesthetics of the wearables, properly motivating a user to alter engrained behaviour is crucial in sustaining long-term engagement. One aspect of sustainable wearable use is by embedding behavioural elements in the design and functionality of the wearable. This could be by designing the wearable to show aspects of emotion and empathy with the user, rather than dynamically calculating data-in and out. Major wearable manufacturers are focusing on designing elegant product with an engaging software, yet they are largely ignoring the principles of motivation and behaviour change. This refers to promoting health behaviours using technique and principles from behavioural theories. Understanding habit formation, social motivation and goal reinforcement will create sustainable wearable devices to promote health and wellness.

This suggests a need for new theoretically-grounded approaches to engage individuals in the analysis of data collected through self-monitoring that can lead to discovery, insights and, as a result, improved health. Theories of behavioural psychology, such as Theory of Planned Behaviour [1], BJ Fogg's Behaviour Model [9] and Reinforcement Theory of Motivation [33], combined with UX design could be adapted to change users perceived behaviour towards wearables. Moreover, working on user's intrinsic and extrinsic motivation and understanding their personality traits and correlation with the wearable can increase their engagement and help them perceive the technology as useful. Wearable developers should integrate engaging techniques and incentives, such as gamification to trigger users into using the device.

*4.0.4 User Expectation.* Wearables focus on data collection and self-monitoring, such as providing data to users about their activity and letting them use it to improve their health or wellness. The findings from our study and the study by Grudin et al., [12] both revealed the changes in user's lifestyle and living condition could be a factor for abandoning the wearable/smartwatch. For example, some sellers in our study reported they posted their activity tracker for sell after they bought a bike or started a scuba diving activity, where in the first case the tracker wasn't tracking bike rides, and the second case the tracker wasn't waterproof. Despite their ongoing interest, users abandoned their wearable due to a change in their lifestyle. Users are often looking for a unique characteristic a wearable could offer, which is misinterpreted by the wearables. Instead, the features offered by the wearables are mismatched with user expectation which results in abandonment. For example, a seller posted "I sell it because it has no GPS and I find it useless". The trackers also fail to provide real value to the user. Users who abandon their devices, often did so because they did not find them useful, got bored, and did not find a behaviour related connection with the device, hence the wearable didn't bring real value to the user. Most wearables incorporate some form of goal-setting and feedback mechanism, notification features, and social sharing, yet wearables are strictly focused on short-term effects and don't consider behavioural change perspective [18]. A study by Fausset et al. [8] examined technology acceptance and adoption of activity monitoring among older adults. The study found participants initial attitudes were positive; but one person stopped using the technologies after a day, another stopped after eight days, and two before the end of the two-weeks period. The participants reported concerns included inaccurate data collected, wasting time, and uncomfortable to wear. In general, studies have found success in short-term usage of activity trackers, but the devices are still ineffective in sustaining long-term use. Many of wearable owners' loose enthusiasm with the device once the novelty of knowing how many steps they've taken wears off. This could be because they are primitive, or too big, and have inefficient battery life.

## 5 LIMITATIONS

Several limitations are associated with this study, partially related to data collection and the analysis. There are few items reported as damaged, this is since we believe seller wants to promote their item to be sold. Moreover, the majority of posts did not highlight motivation, condition, technical limitation behind the abandonment. This can introduce a bias in the data analysis, where the seller mostly describes positive aspects of the technology and minimising negative details, since they're motivated to maximise profits and quickly sell their item. To illustrate, users with minor damages in their items might downplay the damage of the wearable. Posts reporting damaged were only reported by users posted on Kijiji and there was no such report on Gumtree.

The post analysis on Kijiji returned more posts about Apple Watch and Samsung Gear, however we believe this isn't necessarily because more people are abandoning these trackers. One reason could be because of the big market share these brands own. Moreover, posts that didn't describe negative aspects of their experience with the wearable might question their real motive behind the abandonment. In addition, having some unclear and ambiguous posts made it hard to determine the actual motive behind the posts.

Some abandonment cases were perceived as positive abandonment of the device, similarly to the study by Grudin et al. [12], we found some cases where the abandonment was due to sellers desire to upgrade to more advanced trackers, or it was a gift, or even related to size issues. These causes aren't necessarily abandonment of the tracker, since there is no indication of a drop-in user intention to reuse the application. The study has shown other usage patterns and motives for abandoning the wearable. For example, usage duration before abandonment have shown the timeframe users spend before selling the device. Although this is not sufficient to conclude their actual motives for abandonment, however we believe studying the relation between usage time and user decision and link them with the abandonment will help understand usage pattern among different wearable users. This is outside the scope of this study; however, we invite researchers to conduct some studies on this issue. Technology limitations highlighted by users was another interesting pattern about user motives for abandonment. We have obtained both technical (e.g., I sell by switching to 2 series because of the need for GPS) and non-technical (e.g., Purchased for curiosity) indications about the abandonment.

## 6 DISCUSSION

There is a grim picture regarding wearable technologies and their ability to enhance long-term engagement to achieve meaningful goals or enact changes in user's health behaviours [14, 15, 21]. Recent studies documented a high rate wearable abandonment by their users [5, 12]. These studies questioned the core functionality of wearable technologies and concluded that either the overall vision for these technologies is misplaced, their design is deeply flawed, or both. However, our study shows that there exist many reasons for the abandonment beyond the technical and design aspects. User intention and wearable capabilities, flexibility of the wearable to adapt and user habits are all factors affecting user engagement with the wearable in the long-term. Using persuasive elements in the wearable design should clearly define the role of the persuasive technology in sustaining user adherence. For example, whether the persuasive tool will be permanent and always present in the system or it will be temporary and will be off after the user achieves a certain level. In addition, the role of behaviour change theories in wearable technologies often takes a static view of the user and do not account for changes in user's circumstances. Adding more flexibility to the wearables could adapt to user's changing circumstances. There is a need to account for new streams of information, flexibility with user grow, and focus on design aesthetics when making wearable technologies for health. In summary, current research focuses on technical and device related limitations and tries to account for user's low adoption rate by further investigating these points. Fewer studies have focused on user habit and behaviour oriented abandonment cases in wearable and smartwatch technology. Our study was initially based on a study by Clawson et al., [5] and their analysis of wearable posts on Crageslist. Based on our findings, even though we analysed posts from Italy and UK, we found several common usage/abandonment patterns among users obtained from both datasets, which were compatible with the findings by Clawson et al. on Crageslist data. For example, the wearable condition, the abandonment motives, and the adoption method found to be common between our datasets from both Kijiji and Gumtree posts and the dataset from Crageslist analysed by the other study.

## 7 CONCLUSION

There is a big disruption with personal health tracking technologies, as they are rapidly adopted into mainstream culture and have sparked an explosion of interest in tracking various aspects of health. However, these technologies suffer from being largely abandoned in the long-term. Current research investigating this issue focused on technical aspects of the abandonment related to wearable features and functionalities. While this is necessary to improve the quality of services offered. However, behavioural aspect plays a great role in understanding motives behind this abandonment, which was relatively unexplored. In this study, we reviewed 484 posts on Kijiji and 624 posts on Gumtree selling second hand wearables/smartwatches. We investigated the technical limitations of the abandonment; however, our focus was on the behavioural as- pect. For that, an iterative analysis was conducted on all posts and we extracted useful insights and patterns of abandonment from all the posts. The findings revealed cases of abandonment, technology limitations highlighted by users, and user intention to the post announcement. In many cases users abandoned their device for personal reasons, and not necessarily due to technical limitations. Understanding user's circumstances, their intend of use, and what the wearable offers could enhance the design and long-term adop- tion of such technology. We provided a list of insights to consider for further development. These insights suggest more research on design aspect, theoretical foundation for user behaviour, motives and expectations from wearables.